\documentclass[aps,prd,preprint,nofootinbib]{revtex4}
\usepackage{amsfonts}
\usepackage{mathrsfs}
\usepackage{graphicx}
\usepackage{amsmath}
\usepackage{amssymb}
\usepackage{bm}
\usepackage{multirow}
\usepackage{epsfig}
\usepackage{subfig}
\graphicspath{{fig/}{dia/}} \DeclareGraphicsExtensions{.eps}
\newcommand{\bqa}{\begin{eqnarray}}
\newcommand{\eqa}{\end{eqnarray}}
\newcommand{\beq}{\begin{equation}}
\newcommand{\eeq}{\end{equation}}

\begin{document}
\baselineskip 20pt

\title{Exploring Bosonic Mediator of Interaction at BESIII}

\author{Jun Jiang$^{a}$\footnote{jiangjun87@sdu.edu.cn}}
\author{Hao Yang$^{b}$\footnote{yanghao174@mails.ucas.ac.cn}}
\author{Cong-Feng Qiao$^{b,c}$\footnote{qiaocf@ucas.ac.cn, corresponding author}}

\affiliation{$^a$School of Physics, Shandong University, Jinan, Shandong 250100, China\\
$^b$School of Physical Sciences, University of Chinese Academy of Sciences, Beijing 100049, China\\
$^c$CAS Center for Excellence in Particle Physics, Beijing 100049, China}

\author{~\\}

\begin{abstract}

We present a comprehensive investigation on the possibility of the search for new force mediator $X$ boson in $e^+e^-$ collision and $J/\psi$ decay at the BESIII experiment.
The typical interactions of $X$ boson coupling to leptons and quarks are explored.
The production and decay properties of this $X$ particle, the product/decay chains $e^+e^-\to X\gamma \to e^+e^-\gamma$ and $J/\psi \to X\gamma\to\mu^+\mu^-\gamma$, and exclusion limits on the reduced coupling strength parameters as functions of $X$ boson mass are presented.
With the data set of tens of fb$^{-1}~e^+e^-$ or $10^{10}~J/\psi$, we find that the exclusion limits on the coupling strength parameters fall in the range of $10^{-3}\sim10^{-4}$, depending on $m_X$ assuming the decay width 10 eV$<\Gamma_X<$100 eV reasonably, for various hypotheses in the literature.
According to our estimation, the search for new force mediator $X$ boson in both $e^+e^-$ collision and $J/\psi$ decay are accessible in nowadays BESIII experiment.

\vspace {7mm} \noindent {PACS numbers: 14.70.Pw, 13.66.Hk, 13.20.Gd}

\end{abstract}
\maketitle

\section{Introduction}

The strong and electroweak interactions in between ordinary matter are described well by the Standard Model (SM) of particle physics, but new physics must be responsible for the dark matter, the matter-antimatter asymmetry, {\it etc.} \cite{Tanabashi:2018oca}.
It is quite possible that a more complete theory with additional gauge interactions may provide solutions to those mysteries.
This motivates experimental searches for the non-SM gauge bosons, named $X$ in this work, which mediate such extended interactions.
In general, the $X$ particle can be a scalar/pseudoscalr Higgs-like particle, the dark photon or $Z^0$-like particle, {\it etc}.

Searching for the new interaction mediator $X$ is appealing and challenging.
At the Large Hadron Collider, CMS and ATLAS collaborations search for the $Z^\prime$ and $W^\prime$ gauge bosons in the mass range from several hundreds of GeV to several TeV \cite{Aaboud:2018zba,Aaboud:2018jux,Kalsi:2018dgi,CMS:2018jxa}.
The light CP-odd Higgs-like particle is explored at both the B-factories \cite{Ahmed:2016ift,TheBelle:2015mwa} and BESIII experiment \cite{Ablikim:2015voa,Ablikim:2011es}.
For the dark photon, various experiments have been searching for it in a broad mass range, like the NA64 experiment \cite{Banerjee:2018vgk}, BaBar \cite{Lees:2017lec}, LHCb \cite{Aaij:2017rft}, KLOE-II \cite{Bloise:2015wfb}, HADES \cite{Agakishiev:2013fwl}, as well as future experimental projects like PADME \cite{Raggi:2018doy}, VEPP-3 \cite{Wojtsekhowski:2017ijn}, DarkLight \cite{Corliss:2017tms}, {\it etc}.
For the probe of axion-like particles, see \cite{Bauer:2017ris,Bauer:2018uxu} and references therein.

Here, we focus on the search of the new interaction mediator $X$ at the BESIII detector.
The BESIII experiment works in the C.M.S. energy region of $2\sim4.6$ GeV and has accumulated 1.3 billion $J/\psi$'s and 0.5 billion $\psi(3686)$'s \cite{Wang:2018ycf}, which provides ideal samples for the new interaction mediator search in the mass region up to several GeV.\footnote{The BESIII collaboration has announced that they have finished accumulating a sample of 10 billion $J/\psi$ events on February 11, 2019.}
The latest result is the dark photon ($U$) search in $J/\psi\to\eta^{\prime}U$ process followed by $U\to e^+e^-$ decay. The mass range of $0.1\sim2.1$ GeV  was explored but no significant signal observed \cite{Ablikim:2018bhf}.
The initial state radiation (ISR) processes $e^+e^-\to e^+e^-\gamma_{ISR}$ and $e^+e^-\to \mu^+\mu^-\gamma_{ISR}$ were explored in the mass range of 1.5 up to 3.4 GeV, yet no enhancement found in the invariant-mass spectrum of the leptonic pairs \cite{Ablikim:2017aab}.
Using the abundant $\psi^{\prime}$ and $J/\psi$ data sets, the decay chains of $\psi^{\prime} \to J/\psi\pi^+\pi^-, J/\psi \to A^0\gamma, A^0\to\mu^+\mu^-$ \cite{Ablikim:2011es} and $J/\psi \to A^0\gamma, A^0\to\mu^+\mu^-$ \cite{Ablikim:2015voa} were employed to search for the light CP-odd Higgs-like particle $A^0$.

Theoretically, when considering the interactions between the new force mediator $X$ and SM particles, extra Lagrangian needs to be introduced to the SM.
For dark photons, the new coupling arises from the gauge-invariant ``kinetic mixing'' of the new Abelian gauge group $U(1)_X$ with the SM hypercharge group $U(1)_Y$ \cite{Holdom:1986eq,Dienes:1996zr}.
For a $Z^0$-like $X$ boson, the theory with axial-vector couplings can also be UV-completed consistent with the SM gauge invariance \cite{Kahn:2016vjr,Ismail:2016tod}.
And CP-odd pseudoscalar Higgs bosons are suggested in the next-to-minimal super-symmetric standard model \cite{Ellwanger:2009dp}, where the mass of the lightest CP-odd Higgs boson can be less than that of a $J/\psi$.
In fact, many literature try to explore the models beyond SM by adding extra gauge interactions or introducing new gauge symmetries, see \cite{Fayet:1989mq,Das:1999hn,Emam:2007dy,Kyae:2013hda,Dong:2016gxl,Alonso:2017uky,Nomura:2017lsn} as examples.

In the phenomenological study, various possible phenomenons on the new interaction mediator $X$ were investigated.
One latest example, a ``fifth force'' mediated by a protophobic 16.7 MeV boson was suggested to explain the $^8Be^*$ anomaly \cite{Krasznahorkay:2015iga,Feng:2016jff,Gu:2016ege}, which might also be a solution to the muon's anomalous magnetic moment \cite{Feng:2016jff,Gu:2016ege}, NuTeV anomaly \cite{Liang:2016ffe} or 511 keV line \cite{Jia:2017iyc}.
Recently, the NA64 experiment presented the first direct search for this hypothetical 16.7 MeV boson and exclude part of its allowed parameter space \cite{Banerjee:2018vgk}.
For a very long period, various literature are discussing the possibility of the search for the new interaction mediator $X$ in current and future experiments, see Refs. \cite{Fayet:1981rp,Batell:2009yf,Bjorken:2009mm,Li:2009wz,Alikhanov:2017cpy,Araki:2017wyg,Alioli:2017nzr,He:2017zzr} for instance.

In this work, we study the production of the new interaction mediator $X$ boson associated with a photon in both electron-positron collision and $J/\psi$ decay at the BESIII experiment, in which different theoretical hypotheses on the nature of $X$ will be taken into account.
In section II, we explore the mediator $X$ boson in the $e^+e^-\to X\gamma \to e^+e^-\gamma$ product chain. Under the BESIII experiment conditions, the production and decay properties of this new $X$ particle, and the exclusion limits on the reduced coupling strength parameters of $X$ boson to SM particles as functions of the $X$ boson mass will be studied.
In section III, we explore the $X$ boson in the decay chain of $J/\psi\to X\gamma \to \mu^+\mu^-\gamma$.
Section IV is reserved as the summery and outlook.
Finally, some useful formulas are displayed in the Appendix. 
It should be mentioned that several formulas presented in this manuscript about the $Z^0$-like $X$ hypothesis overlap with those in our previous work \cite{Jiang:2018uhs}, however with different concerns.

\section{$X$ Production in $e^+e^-$ Collision}

As an interaction mediator, the new particle $X$ is usually regarded as a boson.
In this section, we study the production and decay properties of $X$ boson in electron-positron collision for both spin-0 and spin-1 hypotheses.
In particular, we will discuss the decay width of the $X$ boson, and the exclusion limits on the reduced coupling strength parameters as functions of $X$ boson mass under the BESIII experiment conditions.

\subsection{Spin-1 Hypothesis}

As a general case, extra Lagrangian of the spin-1 $X$ boson can be formulated as
\bqa
\mathcal{L}_X=-\frac{1}{4}X_{\mu\nu}X^{\mu\nu} + \frac{1}{2}m_X^2X_{\mu}X^{\mu}-\sum_f e\bar{f}\gamma_\mu (\epsilon_v-\epsilon_a\gamma_5) f X^\mu,\label{lag}
\eqa
where $e$ is the electron charge and $\epsilon_{v/a}$ denote the reduced coupling strength of new boson $X$ to vector/axial-vector currents, which implies the $X$ boson can be either the dark photon or a $Z^0$-like particle. In Eq. (\ref{lag}), we simply assume the coupling strengths of the new particle to leptons and quarks are equal.

\begin{figure}[htbp]
\begin{center}
\includegraphics[scale=0.8]{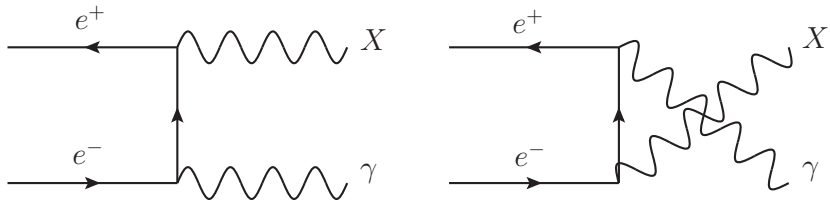}
\caption{The Feynman diagrams of the $e^+e^- \to X+\gamma$ process, where the new particle $X$ can be a massive neutral spin-0 or spin-1 particle.}\label{fd}
\end{center}
\end{figure}

We consider the production of the new boson $X$ in associated with a SM photon in $e^+e^-$ collision, whose Born level Feynman diagrams are displayed in Fig. \ref{fd}.
And the differential cross section of the process $e^+e^- \to X+\gamma$ with respect to cos$\theta$ can be formulated,
\bqa
\frac{d\sigma}{d\cos\theta}=\frac{2\pi \alpha^2(s-m_X^2)}{16s^{3/2}\sqrt{s-4m_e^2}}\big(\epsilon_v^2 |M_v|^2+\epsilon_a^2 |M_a|^2\big), \label{dif-eq}
\eqa
with
\bqa
|M_v|^2&=
\frac{32 s (\cos^2\theta (4 \delta^2 m_X^2-s) (s (4 \delta^2 m_X^2+s)+m_X^4)+s (m_X^4 (1-16\delta^4) +4 \delta^2 m_X^2 (s-2 m_X^2)+s^2))}{(m_X^2-s)^2 (s-\cos^2\theta (s-4 \delta^2 m_X^2))^2}-16,  \label{dif-eq-1} \\
|M_a|^2&=
\frac{32 s (8 (4 \delta^2-1) \delta^2 m_X^4 s+(m_X^4+2 \delta^2 (m_X^4-6 m_X^2 s+s^2)+s^2) (s-\cos^2\theta (s-4 \delta^2 m_X^2)))}{(m_X^2-s)^2 (s-\cos^2\theta (s-4 \delta^2 m_X^2))^2}-16. \label{dif-eq-2}
\eqa
Here, $\delta$ stands for the ratio $m_e/m_X$, $\theta$ is the emitting angle of the photon with respect to the $e^+e^-$ beam axis, $\sqrt{s}$ is the C.M.S energy, $m_{X/e}$ are the masses of the $X$/electrons, and $\alpha=1/137$ is the fine structure constant. In the $\delta\to0$ (or $m_e\to0$) limit, we have
\bqa
|M_v|^2=|M_a|^2=
16(\frac{2(m_X^4+s^2)}{\sin^2\theta(m_X^2-s)^2}-1).
\eqa

\begin{figure}[htbp]
\begin{center}
\includegraphics[scale=0.8]{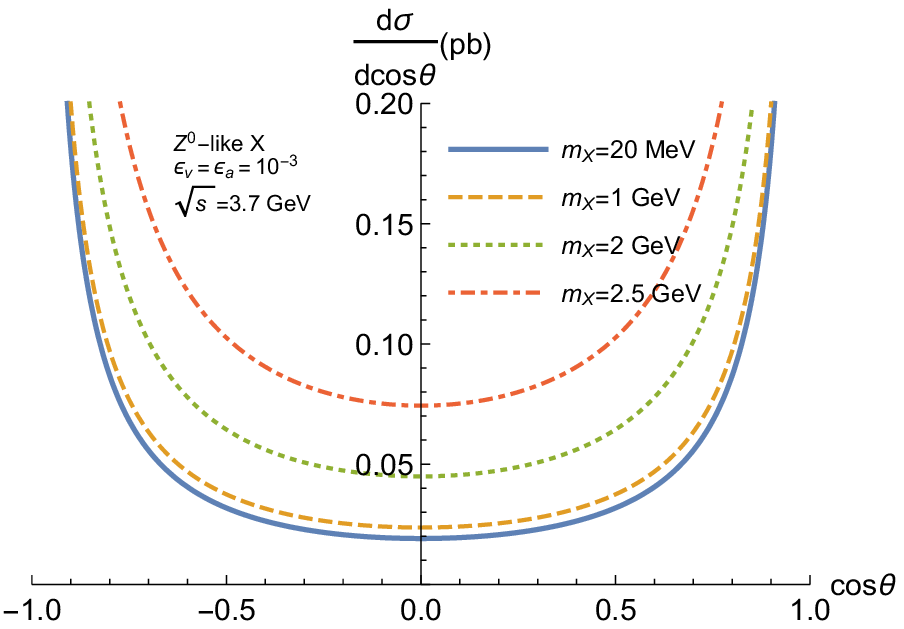}
\includegraphics[scale=0.9]{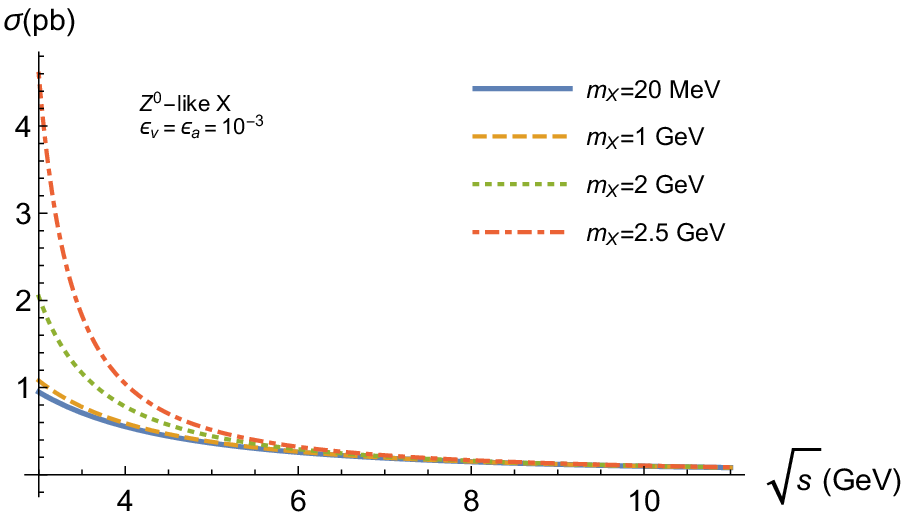}
\caption{The differential cross section with respect to cos$\theta$ of the $e^+e^- \to X+\gamma$ process and its total cross section as a function of $\sqrt{s}$ for the $Z^0$-like $X$.}\label{coss}
\end{center}
\end{figure}

\begin{figure}[htbp]
\begin{center}
\includegraphics[scale=1]{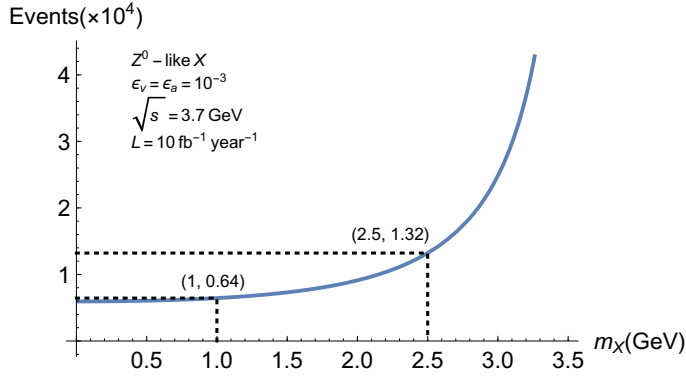}
\caption{The events of the $Z^0$-like $X$ produced per year as a function of its mass $m_X$ through the $e^+e^- \to X+\gamma$ process at $\sqrt{s}$=3.7 GeV. The 93\% solid angle coverage of the BESIII detector is considered.} \label{events}
\end{center}
\end{figure}

In Fig. \ref{coss}, we plot the differential distribution $d\sigma/d$cos$\theta$, and the total cross section as a function of $\sqrt{s}$.
Here we consider four $m_X$ inputs, and adopt the reduced coupling strength parameters $\epsilon_{v}=\epsilon_{a}=10^{-3}$.
Running at $\sqrt{s}=3.7$ GeV, the luminosity of BESIII can reach $10^{33}$ cm$^{-2}$s$^{-1}$$\simeq10$ fb$^{-1}$year$^{-1}$ \cite{Han:2005mu}.
Then we can estimate the events of the $Z^0$-like $X$ boson produced per year as a function of its mass $m_X$ in $e^+e^- \to X+\gamma$ process, as presented in Fig. \ref{events}.
Here we have taken the 93\% solid angle coverage of the BESIII detector into consideration \cite{Ablikim:2009aa}. 
For 1 GeV $<m_X<3.4$ GeV, we can obtain about $(0.64\sim6.5)\times10^4$ $Z^0$-like $X$ bosons. 
And events at $m_X\sim2.5$ GeV are about two times of those where $m_X<1$ GeV.
In Fig.s \ref{coss} and \ref{events}, it is found that these observables are not sensitive to the new boson mass $m_X$ in the region of $m_X<1$ GeV.
Here one can easily estimate the results when adopting other $\epsilon_{v/a}$ inputs, since the (differential) cross sections are proportional to the squared reduced coupling strength parameters $\epsilon^2_{v/a}$.


Experimentally, the new boson $X$ would be reconstructed with its decay products.
Since its mass $m_X$ may range from tens of MeV to several GeV at BESIII, there might be various decay products. 
Below the 2$\pi$ threshold ($\sim$ 270 MeV), the new boson $X$ can decay into $e^+e^-$, $\mu^+\mu^-$, photons, $\nu\bar{\nu}$ or light dark sector particles.
While heavier $X$ boson may have various hadronic decay products.
In this section, we prefer using the $e^+e^-$ pairs to fully reconstruct the new $X$ boson.
And its decay width is
\bqa
\Gamma_0\equiv\Gamma(X\rightarrow e^+e^-)= \frac{\alpha\sqrt{m^2_X-4m_e^2}}{3m^2_X}\big(\epsilon_v^2(m_X^2+2m_e^2)+\epsilon_a^2(m_X^2-4m_e^2)\big).\label{avdw}
\eqa
For $m_X=$ 16.7 MeV, 212 MeV, 1 GeV and 3.4 GeV, we obtain the decay width $\Gamma_0 =$ 0.081, 1.0, 4.9 and 17 eV respectively by assuming $\epsilon_v=\epsilon_a=10^{-3}$.
Note that, when reconstructing the $X$ boson signals in the invariant-mass spectrum of the $e^+e^-$ pairs ($M_{ee}$), the smaller the decay width $\Gamma_0$ is, the more clear the $X$ resonance bump will be.
Below the 2$\mu$ threshold ($\sim$ 212 MeV), since the 2$\gamma/3\gamma$ decay modes are highly suppressed, and assuming reasonably the decay width of $X\to \nu\bar{\nu}$ approximately equals to $\Gamma_0$, we can estimate the total decay width $\Gamma_X$ of $X$ boson would be four times of $\Gamma_0$, {\it i.e.} $\Gamma_X\lesssim4$ eV.
While for heavier $X$ boson whose mass is above the $2K$ threshold ($\sim$ 1 GeV), we can roughly estimate the hadronic decay width is two times of $\Gamma_0$, and thus the total decay width $\Gamma_X$ would be seven times of that, {\it i.e.} $\Gamma_X$ lies in $34\sim119$ eV for 1 GeV $<m_X<3.4$ GeV.\footnote{$\sum\limits_{q=u,d,s} \Gamma(X \to q\bar{q})/\Gamma_0\approx2$, $\Gamma(X\to \mu^+\mu^-)/\Gamma_0\approx1$. Besides, for $m_X$ lies between the 2$\tau$ threshold ($\sim$3.45 GeV) and $\sqrt{s}=3.7$ GeV, the $X \to \tau^+\tau^-$ decay mode will contribute.}
It is worth noting that the total decay width $\Gamma_X$ might be much bigger than above estimation, since the interaction mediator $X$ may decay into some dark sector particles (dark decay width) where the coupling strength can be out of the $\epsilon_{v/a}^2$ suppression.
In the follow analysis, we will adopt the decay width $\Gamma_X=10$ eV, 100 eV and 1 keV as three reasonable parameter options.\footnote{When not considering the dark decay width, we obtain $\Gamma_X=10$ eV at $m_X=0.31$ GeV, and $\Gamma_X=100$ eV at $m_X=2.94$ GeV, and this mass span is covered by the BESIII energy region.}

In the laboratory frame of $e^+e^- \to X+\gamma$ process, the velocity of $X$ boson is $v = \frac{E_{\gamma}} {\sqrt{E_{\gamma}^2+m_X^2}}$ with the energy of emitting photon $E_{\gamma}=\frac{s-m_X^2}{2\sqrt{s}}$.
Then one can estimate the decay length of $X$ boson decaying into $e^+e^-$ pair as $\frac{1}{\Gamma_0}\times\frac{\hbar c}{\sqrt{1-v^2}}$, where $\hbar$ and $c$ are the reduced Planck constant and velocity of light respectively.
In the $m_e\to 0$ limit and assuming $\epsilon_v=\epsilon_a=\epsilon$, the decay length turns into $\frac{3\hbar c}{\alpha m_X\epsilon^2\sqrt{1-v^2}}$.
By setting $\epsilon=10^{-3}$ and $\sqrt{s}=3.7$ GeV, the decay length would be about 0.012 $\sim$ 1.7 $\mu$m for 212 MeV $<m_X<3.4$ GeV.
It is meaningful to measure the decay length of $X$ boson since it can help to determine the coupling strength and to identify signals over the background.

For the dark photon hypothesis, one can obtain the cross section or decay width by setting $\epsilon_a=0$ in the relevant formulas above.
We find that numerical values of the differential distribution $d\sigma/d\cos\theta$, total cross section as a function of $\sqrt{s}$, the events as a function of $m_X$ and the decay width $\Gamma_0$ for the dark photon $X$ boson are one half of those for the $Z^0$-like $X$ boson, {\it i.e.} the contributions of vector and axial-vector currents are of the equal importance.

\begin{figure}[t]
\begin{center}
\includegraphics[scale=0.48]{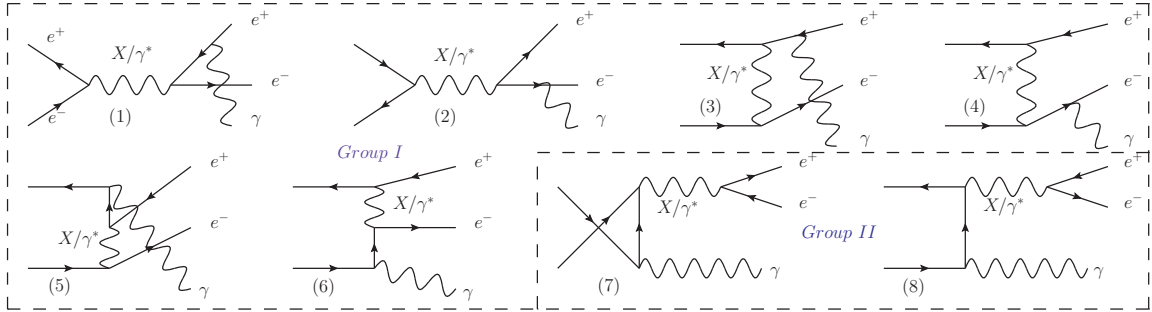}
\caption{The Feynman diagrams of the process $e^+e^- \to e^+e^-\gamma$, where both the $X$ propagated diagrams (the signal) and the virtual photon propagated ones (the background) are displayed. Diagrams are divided into two groups (I, II).} \label{eeg}
\end{center}
\end{figure}

Next we explore the mediator $X$ boson in the process of $e^+e^- \to e^+e^-\gamma$, and the Born level Feynman diagrams are displayed in Fig. \ref{eeg}, where the propagator can be either a $X$ boson (the signal) or a virtual photon (the background).
To be noted that the last two signal diagrams (Fig. \ref{eeg} (7,~8)) are the $X$ resonant ones in the invariant-mass spectrum $M_{ee}$, {\it i.e.} the new boson $X$ and a photon are firstly produced then followed by the $X$ to $e^+e^-$ decay.
For the sake of discussion, these two diagrams are grouped in (II), while the previous six are grouped in (I).
In the calculation of the group (II), we adopt the Breit-Wigner form for the propagator of the unstable $X$ boson, \emph{i.e.} $1/(p_X^2-m_X^2+i\Gamma_Xm_X)$, where $p_X^2=M_{ee}^2$ and $\Gamma_X$ is the total decay width of $X$ boson.
In our numerical calculation, we find that the group (II) dominates the signal process for both spin-1 and spin-0 $X$ boson hypotheses.
So we take only the group (II) of the $X$-propagated diagrams into consideration in the following discussion.

\begin{figure}[htbp]
\begin{center}
\includegraphics[scale=1]{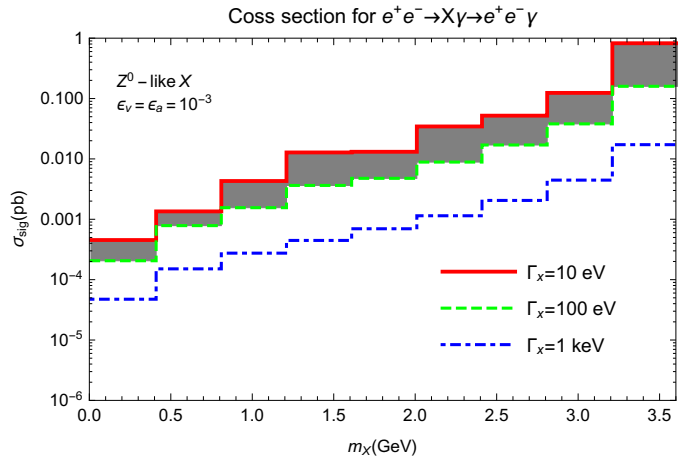}
\caption{The cross section $\sigma_{sig}$ of the signal process $e^+e^- \to X \gamma \to e^+e^-\gamma$ as a function of $X$ boson mass $m_X$ for $Z^0$-like $X$ boson. The total decay width of $X$ boson varies in the range of $\Gamma_X=10\sim100$ eV if its mass lies in $m_X=0.31\sim2.94$ GeV. And the $\Gamma_X=1$ keV curve is for the large dark decay width hypothesis.}\label{sigcsav}
\end{center}
\end{figure}

Under the BESIII experiment conditions\footnote{We adopt the selection conditions for the tracks in final states, {\it i.e.} the photon selection condition is $|\cos\alpha|<0.8$ with its energy $E_{\gamma}>25~\text{MeV}$ for the barrel (omitting the narrow endcap region), while good charged tracks are constrained in the region of $|\cos\beta|<0.93$, with $\alpha/\beta$ being the polar angles of the final particles with respect to the $e^+e^-$ beam axis \cite{Ablikim:2009aa}.}, we evaluate the cross section $\sigma_{sig}$ of the signal process $e^+e^- \to X \gamma \to e^+e^-\gamma$ as a function of the $X$ boson mass $m_X$ for the $Z^0$-like $X$, which is presented in Fig. \ref{sigcsav}.
Since the total decay width of $X$ boson varies in the range of $\Gamma_X=10\sim100$ eV if its mass lies in $m_X=0.31\sim2.94$ GeV, we marked the corresponding area in the plot. And the $\Gamma_X=1$ keV curve corresponds with the large dark decay width hypothesis.
It is found that the cross sections increase with the decrease of the $\Gamma_X$. This can be attributed to the dominated Feynman diagrams of the group (II), since a smaller $\Gamma_X$ in the Breit-Wigner propagator will lead to a larger cross section around the $X$ resonance point of $M_{ee}=m_X$.
For the dark photon $X$ boson, we find that its cross section is about one fourth of the $Z^0$-like $X$ one, {\it i.e.} $\sigma_{Z^0-\text{like}~X}/\sigma_{\text{dark photon}~X}\approx4$, which implies again that the axial-vector current and the vector one are of the same importance.
For the background process displayed in Fig. \ref{eeg} ($\gamma^*$-propagated diagrams), we need to consider only the initial state radiation (ISR) diagrams of Fig.s \ref{eeg} (5$\sim$8). The reason is that the ISR $X$-propagated Fig.s \ref{eeg} (7,~8) dominate the signal process, thus we can study only the ISR processes in experiments. We obtain the cross sections for the background $\sigma_{bg}=13$ nb for the ISR diagrams only, and 35 nb for all $\gamma^*$-propagated ones.

For a sound estimation, we need to know the signal-to-noise ratios (SNR), which would judge whether we can identify the signals over the background or not. 
Taking the $\Gamma_X=10$ eV case as an example, we select the central value in each $m_X$ bin of Fig. \ref{sigcsav}, and evaluate the cross section of the background process and the SNR around the signal bump in the $M_{ee}$ spectrum (assuming $\epsilon_v=\epsilon_a=\epsilon$). 
Because the BESIII detector has the limited energy resolution ($\delta_E=2.5\%\sqrt{E}$ around 1 GeV \cite{Asner:2008nq}), the reconstructed signal shape in the $M_{ee}$ spectrum would be a bump with some width rather than a sharp peak. 
And for a normal distribution, three standard deviations could cover 99.7\% data. 
So we would estimate the cross section of the background process in a $M_{ee}$ span around the selected $m_X$ value with the span width of three standard deviations of the energy resolution, and then calculate the SNR in those spans for the $Z^0$-like $X$ boson, as presented in Tab. \ref{snrav}. Note that only the contribution of the ISR background process is considered. For more analysis on the signal reconstruction and SNR estimation, please refer to our previous work \cite{Jiang:2018uhs}.

\begin{center}
	\begin{table}[htbp]
		\caption{The cross section of the background process in a $M_{ee}$ span around the selected $m_X$ value with the span width of three standard deviations of the energy resolution ($\delta_E=25$ MeV), and the signal-to-noise ratios (SNR) in those spans for the $Z^0$-like $X$ boson (assuming $\epsilon_v=\epsilon_a=\epsilon$).}\label{snrav}
		\begin{tabular}{|c| c|| c| c|| c| }
			\hline
			\multicolumn{2}{|c||}{Signal process} & \multicolumn{2}{|c||}{Background process} & \multirow{2}*{SNR ($\frac{\sigma_{sig}}{\sqrt{\sigma_{sig}+\sigma_{bg}}}$)} \\ \cline{1-4} 
			  $m_{X}$ (GeV) & $\sigma_{sig}$ (pb) ($\Gamma_X=10~\text{eV},\epsilon=10^{-3}$) & $M_{ee}$ (GeV) & $\sigma_{bg}$ (pb) (ISR only)& \\ \hline
			 0.2 & 0.00045 & [0.125, 0.275]  & 42 & 0.000069\\ \hline
			 0.6 & 0.0014 & [0.525, 0.675] & 32 & 0.00025\\ \hline
			 1.0 & 0.0043 & [0.925, 1.075] & 36 & 0.00071\\ \hline
			 1.4 & 0.013 & [1.325, 1.475] & 45 & 0.0019\\ \hline
			 1.8 & 0.013 & [1.725, 1.875] & 61 & 0.0017\\ \hline
			 2.2 & 0.034 & [2.125, 2.275] & 95 & 0.0035\\ \hline
			 2.6 & 0.052 & [2.525, 2.675] & 163 & 0.0041\\ \hline
			 3.0 & 0.12 & [2.925, 3.075] & 379 & 0.0062\\ \hline
			 3.4 & 0.82 & [3.325, 3.475] & 1500 & 0.021\\ \hline
			\hline
		\end{tabular}
	\end{table}
\end{center}

\begin{figure}[htbp]
\begin{center}
\includegraphics[scale=1]{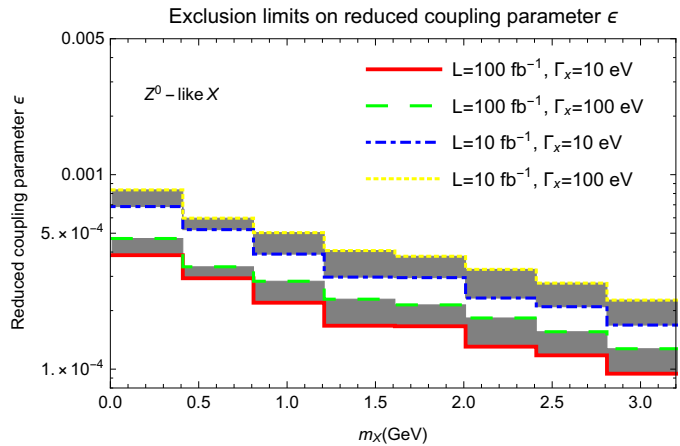}
\caption{The exclusion limit on the reduced coupling strength parameter $\epsilon$ ($\epsilon_v=\epsilon_a=\epsilon$) as a function of the mass $m_X$ for $Z^0$-like $X$ boson. When $m_X$ lies in $m_X=0.31\sim2.94$ GeV, its total decay width would be $\Gamma_X=10\sim100$ eV, {\it i.e.} the shaded areas.} \label{sigexav}
\end{center}
\end{figure}

Finally, we discuss the exclusion limits of the reduced coupling parameter (assuming $\epsilon_v=\epsilon_a=\epsilon$) versus the new boson mass $m_X$ under the BESIII experiment conditions.
In Fig. \ref{sigexav}, regions of the parameter space ($\epsilon$ vs $m_X$) for the $Z^0$-like $X$ boson are presented.
The upper limits on the parameter $\epsilon$ are derived by solving the function, $\sigma_{sig}\times L =1~X$ event, where $\sigma_{sig}$ is proportional to $\epsilon^4$ and $L$ is the integrated luminosity.
With the data of tens of $\text{fb}^{-1}$, it is found that the reduced coupling strength parameter $\epsilon$ is placed in the range between $10^{-3}$ and $10^{-4}$ depending on $m_X$ where $\Gamma_X$ runs from 10 to 100 eV.
For the same reason with the cross section, it is clear that the exclusion limits are sensitive to the decay width $\Gamma_X$.
For the dark photon $X$ boson, the exclusion limits would be suppressed by a factor of $\sqrt{2}$ under the same conditions.

\subsection{Spin-0 Hypothesis}

In this subsection, we will extend our analysis and explore the neutral spin-0 $X$ particle.
We introduce the Yukawa interaction Lagrangians for the scalar ($S$) and pseudoscalar ($PS$) $X$ boson respectively,
\bqa
\mathcal{L}_{S}^{int}=- e\sum_f \eta_f \bar{f} X f,~~~~\mathcal{L}_{PS}^{int}=- e\sum_f \xi_f \bar{f}\gamma_5 X f. \label{lag2}
\eqa
Here $e$ is the electron charge, $\eta_f$ is the reduced Yukawa coupling strength parameter of the scalar $X$ to fermions, and $\xi_f$ stands for that of the pseudoscalar $X$ to fermions.

For the scalar/pseudoscalar $X$ bosons, we also consider their production in the process of $e^+e^- \to X+\gamma$. We obtain the differential cross sections for the scalar/pseudoscalar $X$ with respect to cos$\theta$ as follows,
\bqa
\frac{d\sigma_{S/PS}}{d\cos\theta}=\frac{2\pi \alpha^2(s-m_X^2)}{16s^{3/2}\sqrt{s-4m_e^2}}\times|M_{S/PS}|^2, \label{dif-eq2}
\eqa
with
\bqa
&|M_{S}|^2=
\frac{\eta_e^2 16 s (s (32 \delta^4 m_X^4-8 \delta^2 m_X^2 (m_X^2+s)+m_X^4+s^2)-\cos^2\theta (s-4 \delta^2 m_X^2) (s (s-8 \delta^2 m_X^2)+m_X^4))}{(m_X^2-s)^2 (\cos^2\theta (4 \delta^2 m_X^2-s)+s)^2}, \label{dif-eq2-1}\\
&|M_{PS}|^2=
\frac{\xi_e^2 16 s (s (-8 \delta^2 m_X^2 m_X^2+m_X^4+s^2)+\cos^2\theta (4 \delta^2 m_X^2-s) (m_X^4+s^2))}{(m_X^2-s)^2 (\cos^2\theta (4 \delta^2 m_X^2-s)+s)^2}, \label{dif-eq2-2}
\eqa
where $\eta_e/\xi_e$ are the reduced coupling strength parameters of scalar/pseudoscalar $X$ bosons to electrons, and other variables have the same meanings as those in Eq.s (\ref{dif-eq-1}, \ref{dif-eq-2}). In the $\delta\to0$ (or $m_e\to0$) limit, we have
\bqa
|M_{S}|^2/\eta_e^2=|M_{PS}|^2/\xi_e^2=
\frac{16(m_X^4+s^2)}{\sin^2\theta(m_X^2-s)^2}.
\eqa

For the scalar/pseudoscalar $X$ bosons decaying to the fermion pairs $X\to f\bar{f}$, it is readily to obtain the decay widths,
\bqa
\Gamma(X\rightarrow f\bar{f})_{S}=& \frac{\eta_f^2 C_A \alpha(m^2_X-4m_f^2)^{3/2}}{2 m_X^2},\label{sdw-1}\\
\Gamma(X\rightarrow f\bar{f})_{PS}=& \frac{\xi_f^2 C_A \alpha\sqrt{m^2_X-4m_f^2}}{2}.\label{sdw-2}
\eqa
Here, $C_A= 3$ for quarks and $C_A= 1$ for leptons. Obviously, in the $m_f\ll m_X$ limit, the above two formulas become exactly the same regardless of the coupling parameters.
We find that the decay widths of $\Gamma_0\equiv\Gamma(X\to e^+e^-)$ are numerically equal for scalar and pseudoscalar $X$ bosons when $m_X\ge16.7$ MeV, which are (0.061, 0.99, 3.6, 12.4)$\times10^{-4}$ eV for $m_X=$ 16.7 MeV, 270 MeV ($\sim 2 m_{\pi}$), 1 GeV ($\sim 2 m_K$) and 3.4 GeV respectively by setting $\eta_e=\xi_e=10^{-5}$.\footnote{In SM, the Yukawa coupling strength of the Higgs boson to electrons is $\frac{m_e e}{2m_W sin\theta_w} \approx 6.6 \times 10^{-6}e$.}
Then for $2m_{\pi}<m_X<2m_K$, we can roughly estimate the total decay width $\Gamma_X\simeq\sum\limits_{f=e,\mu,u,d}\Gamma(X\to f\bar{f})=(\frac{m_e^2+m_{\mu}^2+3m_{u}^2+3m_{d}^2}{m_{e}^2})\Gamma_0\approx4.3\times10^4~\Gamma_0\in[4.2,~15.7]$ eV.
And for $2m_K<m_X<2m_{\tau}$, the total decay width would be about $1.5\times10^5~\Gamma_0\in[55,~186]$ eV.
Here we assume the loop induced decay mode $X \to 2\gamma$ is suppressed.

Now we consider the decay length of this spin-0 $X$ boson in the laboratory frame of $e^+e^- \to X+\gamma$ process. Similar to the $Z^0$-like $X$ case, in the $m_e\ll m_X$ limit, the decay length of $X$ boson decaying into $e^+e^-$ pairs is $\frac{2\hbar c}{\alpha m_X\eta_e^2\sqrt{1-v^2}}$, two thirds of the decay length of the $Z^0$-like $X$ regardless of the coupling parameter. By setting $\eta_e=10^{-5}$ and $\sqrt{s}=3.7$ GeV, the decay length would be about 0.16 $\sim$ 22 mm for 212 MeV $<m_X<3.4$ GeV. In the vertex chamber, such a long decay length can be easily observed.

It's worth noting that, in the coming numerical evaluation, we find that all the curves of the scalar $X$ boson hypothesis are overlapped with those of the pseudoscalar $X$ one, including the plot of exclusion limits on the reduced coupling parameter versus the mass $m_X$.
So we only discuss the scalar $X$ case in this subsection.

\begin{figure}[htbp]
\begin{center}
\includegraphics[scale=0.8]{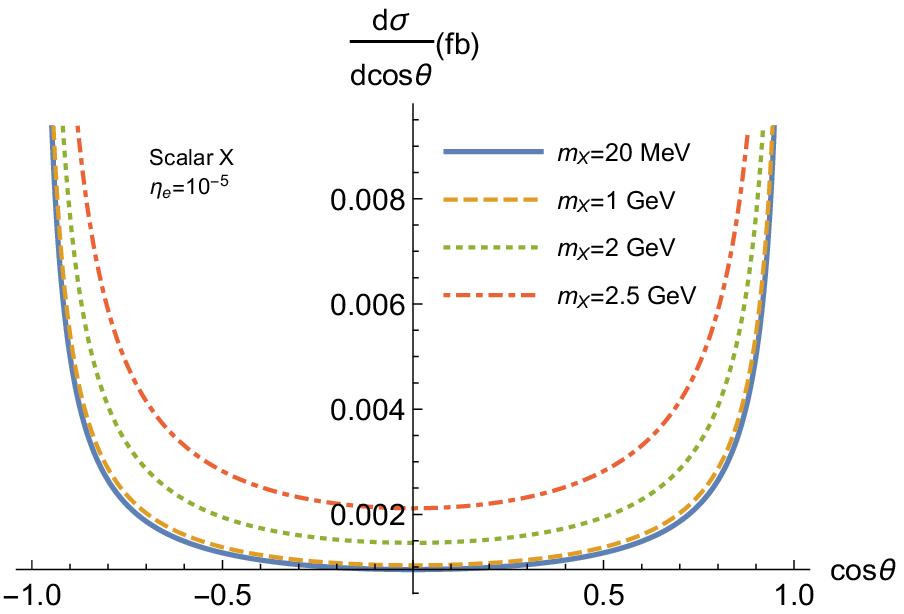}
\includegraphics[scale=0.9]{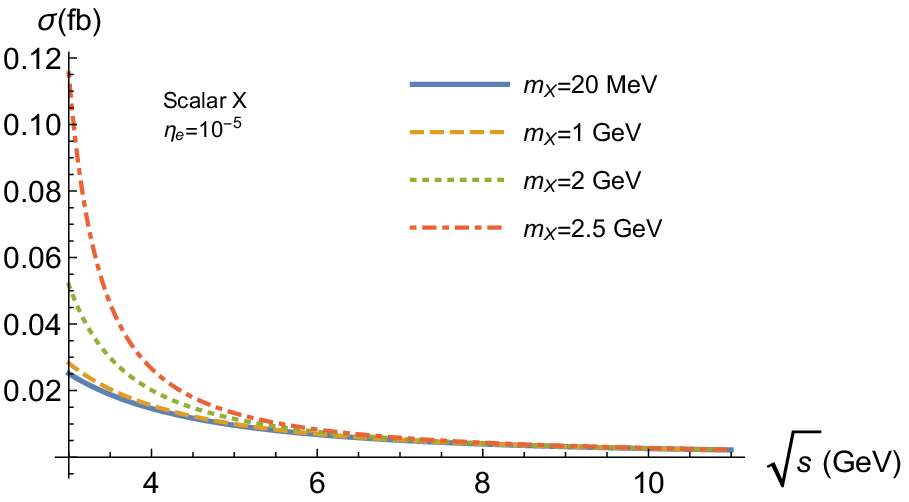}
\caption{The differential cross section of the $e^+e^- \to X+\gamma$ process with respect to cos$\theta$, and the total cross section as a function of $\sqrt{s}$ for the scalar $X$ boson.}\label{scalarcoss}
\end{center}
\end{figure}

\begin{figure}[htbp]
\begin{center}
\includegraphics[scale=1]{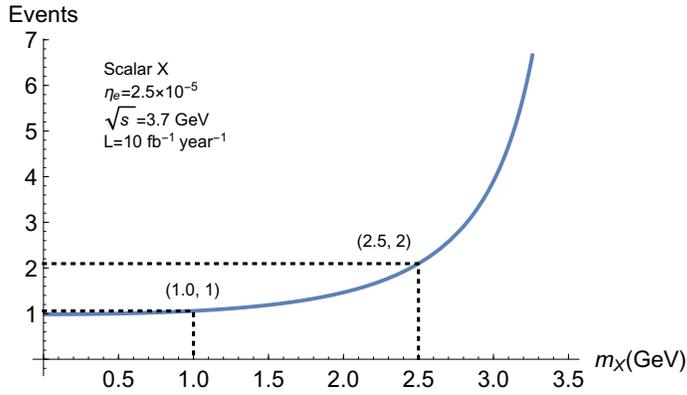}
\caption{The events of the scalar $X$ boson produced per year as functions of its mass $m_X$ in the $e^+e^- \to X+\gamma$ process at $\sqrt{s}$=3.7 GeV. The 93\% solid angle coverage of the BESIII detector is considered.}\label{scalarevents}
\end{center}
\end{figure}

With the help of Eq.s (\ref{dif-eq2}, \ref{dif-eq2-1}), we can evaluate the differential distribution of the cross section $d\Gamma/d$cos$\theta$ and the total cross section as a function of $\sqrt{s}$ for the scalar $X$ boson, which are presented in Fig. \ref{scalarcoss}.
Assuming the reduced Yukawa coupling parameter $\eta_e=2.5\times10^{-5}$ and adopting the luminosity of $L\simeq10$ fb$^{-1}$year$^{-1}$ at $\sqrt{s}=3.7$ GeV, one can also estimate the events of scalar $X$ per year at BESIII detector as displayed in Fig. \ref{scalarevents}.
One may notice that there will be no signals of the scalar $X$ boson for $\eta_e\lesssim2.5\times10^{-5}$ with the $10$ fb$^{-1}$ data set.
In these figures, we find that the values are insensitive to the $X$ boson mass $m_X$ in the region of $m_X<1$ GeV, which is similar to the spin-1 case.
Here one can easily evaluate all the observables when adopting other $\eta_e$ inputs, since the (differential) cross sections are proportional to the squared coupling parameter $\eta_e^2$.


\begin{figure}[htbp]
\begin{center}
\includegraphics[scale=0.85]{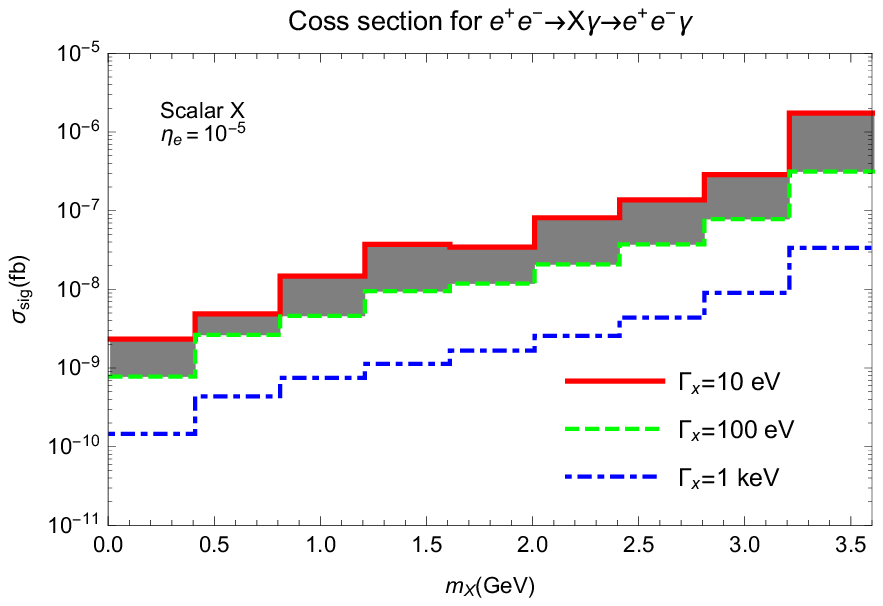}
\includegraphics[scale=0.85]{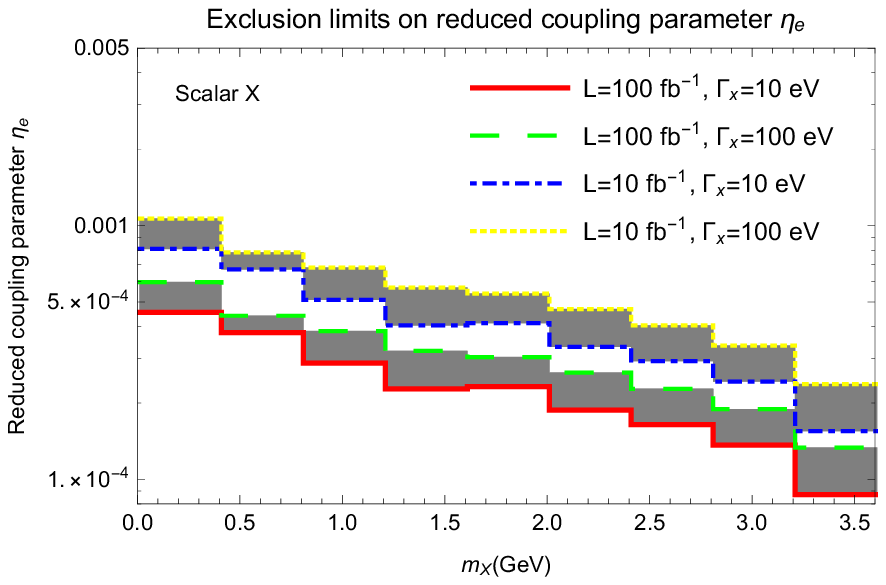}
\caption{The cross section $\sigma_{sig}$ of the signal process and the exclusion limit on the reduced coupling strength parameter $\eta_e$ as functions of $m_X$ for scalar $X$ boson.}\label{sigs}
\end{center}
\end{figure}

For the detection of the scalar $X$ at BESIII, we also use the $e^+e^-$ pairs to fully reconstruct the signals, and both the signal and background Feynman diagrams are the same as those of spin-1 case displayed in Fig. \ref{eeg}.
Here, we continue to adopt the decay width $\Gamma_X=10$ eV, 100 eV and 1 keV as three reasonable parameter options.\footnote{When not considering the dark decay width, we obtain $\Gamma_X=10$ eV at $m_X=0.64$ GeV, and $\Gamma_X=100$ eV at $m_X=1.8$ GeV for this spin-0 $X$ boson.}
Then we evaluate the cross section of the signal process $\sigma_{sig}$ as a function of the $X$ boson mass $m_X$, as is presented in Fig. \ref{sigs}.
Due to the weak Yukawa coupling of $X$ boson to electrons, the cross section is found to be very small.
Similar to the SNR analysis for $Z^0$-like $X$ boson in Tab. \ref{snrav}, we can also evaluate the SNR for the scalar $X$ boson, and find that the SNR for the nine $m_X$ values selected in Tab. \ref{snrav} are all below $10^{-10}$ assuming $\eta_e=10^{-5}$ for $\Gamma_X=10$ eV case.
In Fig. \ref{sigs}, we also present the exclusion limit on the reduced coupling strength parameter $\eta_e$ versus the new boson mass $m_X$.
One can find that the upper limits on the parameter $\eta_e$ lie in the range between $10^{-3}$ and $10^{-4}$ depending on $m_X$ and $\Gamma_X$ for tens of $\text{fb}^{-1}$, which is similar to the spin-1 case.
Note that the BESIII experiment selection conditions are also adopted here.
It is found that the weak Yukawa coupling of $X$ boson to electrons $\eta_e$ exceeds the scope of the upper limits.
In the coming section, we will consider the search for these scalar/pseudoscalar Higgs-like $X$ bosons in a much better process of $J/\psi\to \mu^+\mu^-\gamma$, since the Yukawa coupling strength of $X$ to charm quarks and muons should be much bigger.

As a final remark in this section, the BESIII experiment searched for the $X$ boson in the initial state radiation reactions $e^+e^-\to l^+l^-\gamma_{ISR}$ ($l=e,~\mu)$ using a data set of $2.93$ fb$^{-1}$ at $\sqrt{s}=3.773$ GeV in 2017 \cite{Ablikim:2017aab}.
But no enhancement is observed in the mass range of 1.5 up to 3.4 GeV, the upper limits on the reduced coupling parameters are set to be around $10^{-3}$ with a confidence level of 90\%, which is consistent with our analysis.

\section{$X$ Production in $J/\psi$ Decay}

As is known that the BESIII experiment has accumulated the largest $J/\psi$ data worldwide, and a data set of 10 billion $J/\psi$ events has been obtained in Feb. 11, 2019.
So in this section we will search for the interaction mediator $X$ boson in the $J/\psi \to \mu^+\mu^-\gamma$ decay.
Here, we use the $\mu^+\mu^-$ invariant-mass spectrum rather than the $e^+e^-$ one to reconstruct the $X$ signals, since the Yukawa coupling strength parameters  of $X$ to muons should be much bigger than those of $X$ to electrons in the scalar/pseudoscalar hypotheses.

\subsection{Scalar and Pseudoscalar Hypotheses}

We firstly discuss the production of the scalar($S$)/pseudoscalar($PS$) $X$ bosons in $J/\psi$ decay.
The Yukawa interaction Lagrangian are the same as those of Eq. (\ref{lag2}).
Then we obtain the decay widths of the $J/\psi \to X+\gamma$ process for the scalar and pseudoscalar $X$ bosons respectively,
\bqa
&&\Gamma(J/\psi\to X\gamma)_{S}=\frac{(\eta_c)^2 8 \pi \alpha ^2 \Psi^2 (16 \kappa^4+16 \kappa^2+1)}{27 \kappa^4 (4 \kappa^2-1)m_X^2}\ ,\label{G-s/ps-1}\\
&&\Gamma(J/\psi\to X\gamma)_{PS}=\frac{(\xi_c)^2 8 \pi \alpha ^2 \Psi^2 (4 \kappa^2 - 1)}{27 \kappa^4 m_X^2 }\ , \label{G-s/ps-2}
\eqa
where $\eta_c/\xi_c$ are the reduced coupling strength of scalar/pseudoscalar $X$ to charm quark, $\kappa$ stands for the ratio $m_c/m_X$, $m_c=1.5$ GeV and the squared wave function at the origin $\Psi^2=\frac{m_c^2\Gamma(J/\psi\to e^+e^-)}{4\pi e_c^2 \alpha^2 (1-8\alpha_s/(3\pi))^2}$ \cite{Braaten:2002fi}, with $e_c=2/3$, $\alpha_s=0.23$ and $\Gamma(J/\psi\to e^+e^-)=5.55\times10^{-6}$ GeV \cite{Tanabashi:2018oca}.
Given the total decay width of $J/\psi$ as $\Gamma_{J/\psi}=$ 92.9 keV and the data set of 10 billion $J/\psi$ events, we can estimate the events of the scalar $X$ boson as $(0.63,~1.0,~3.3,~8.6) \times10^9\eta_c^2$, and events of the pseudoscalar $X$ boson as $(0.61,~0.55,~0.34,~0.19) \times10^9\xi_c^2$, for $m_X= (0.212,~1,~2,~2.5)$ GeV accordingly.
If assuming $\eta_c\sim\xi_c\sim10^{-3}$,\footnote{Here we adopt a smaller Yukawa coupling strength in comparison with the one in SM, $\frac{m_c e}{2m_W sin\theta_w}\approx1.9\times10^{-2}e$. } one may obtain $10^2\sim10^3$ events of the scalar/pseudoscalar $X$ bosons.

\begin{figure}[htbp]
\begin{center}
\includegraphics[scale=0.61]{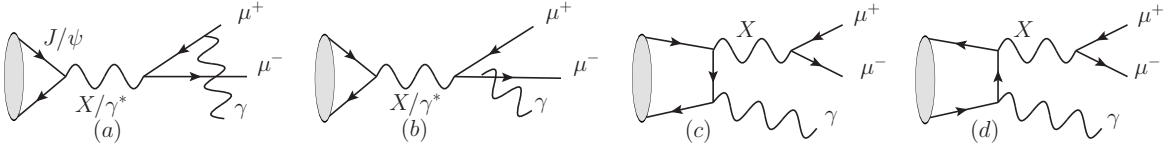}
\caption{The Feynman diagrams of the process $J/\psi \to \mu^+\mu^-\gamma$, where both the $X$ propagated diagrams and the virtual photon propagated ones are displayed.} \label{psimmg}
\end{center}
\end{figure}

For the detection of $X$ boson, it is a better choice to use the $\mu^+\mu^-$ final states rather than the $e^+e^-$ pairs to reconstruct the $X$ signals in the Yukawa coupling hypothesis.
The Born level Feynman diagrams of the $J/\psi(p) \to \mu^+(k_1)+\mu^-(k_2)+\gamma(k_3)$ process mediated by both the $X$ boson (the signal) and the virtual photon (the background) are displayed in Fig. \ref{psimmg}.
One may notice that, because of the conservation of the orbital angular momentum, the background process has only two diagrams of Fig. \ref{psimmg} (a,~b), while only Fig. \ref{psimmg} (c,~d) contribute to the $X$ production in this spin-0 hypothesis.
Here, we also adopt the Breit-Wigner form for the propagators in the calculation of Fig. \ref{psimmg} (c,~d). Then we calculate the differential decay widths $d\Gamma/(d s_1d s_2)$ for the scalar and pseudoscalar $X$ bosons respectively,
\bqa
\frac{d\Gamma_{S/PS}}{d s_1 d s_2}=\frac{\alpha^3 \Psi^2}{768 m_c^4}\times |M_{S/PS}|^2\ ,\label{dfG}
\eqa
with
\bqa
&&|M_{S}|^2=
\frac{(\eta_c\eta_{\mu})^2 1024 (16 m_c^4+16 m_c^2 s_1+s_1^2) (s_1-4 m_{\mu}^2)}{9 (s_1-4 m_c^2)^2 (\Gamma_X^2 m_X^2+(m_X^2-s_1)^2)},  \\
&&|M_{PS}|^2=
\frac{(\xi_c\xi_{\mu})^2 1024 s_1}{9 (\Gamma_X^2 m_X^2+(m_X^2-s_1)^2)}, 
\eqa
where $\eta_{\mu}/\xi_{\mu}$ are the reduced coupling strength parameters of scalar/pseudoscalar $X$ to the muon, $s_1$ and $s_2$ are the Dalitz invariants $s_1=(k_1+k_2)^2$ and $s_2=(k_2+k_3)^2$, and $m_{\mu}$ is the mass of the muon. 

\begin{figure}[htbp]
	\begin{center}
		\includegraphics[scale=0.85]{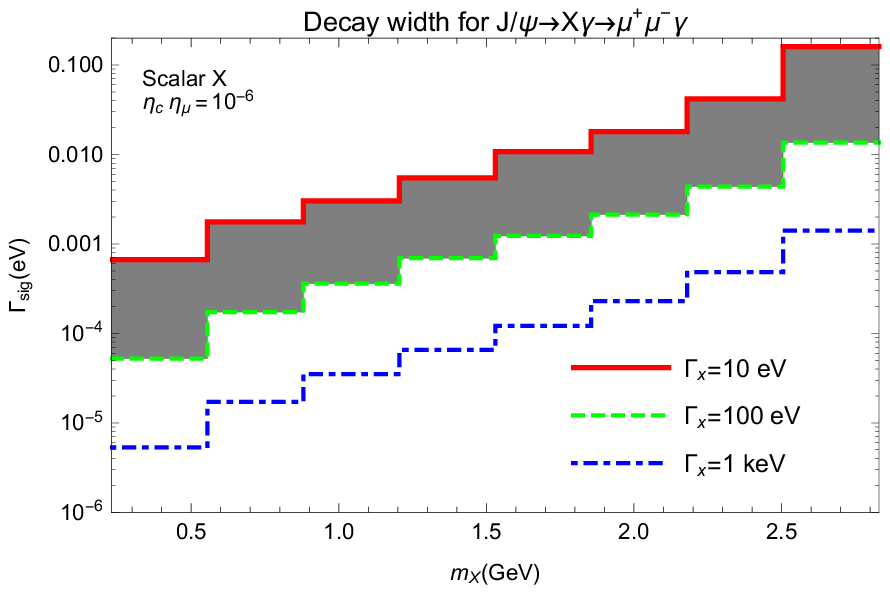}
		\includegraphics[scale=0.85]{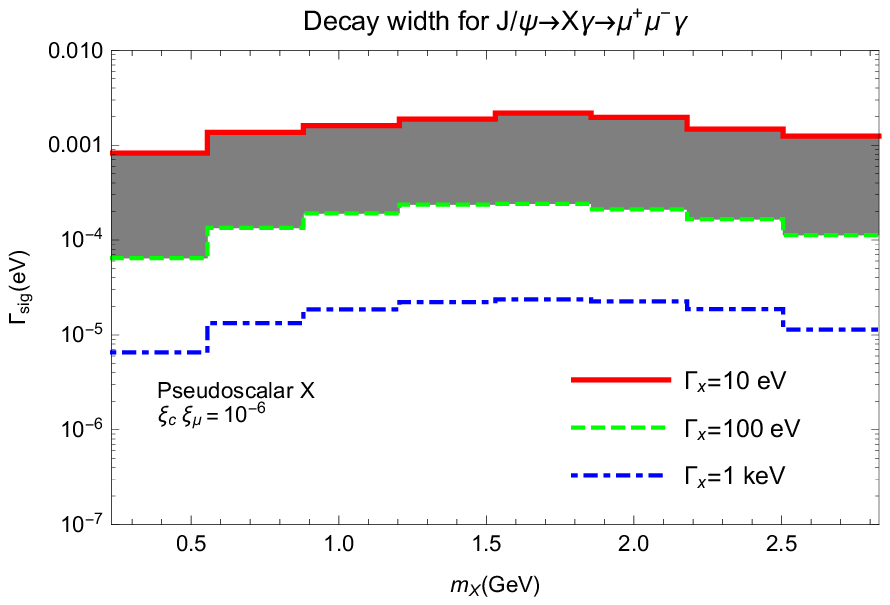}
		\caption{The decay widths $\Gamma_{sig}$ of the signal process $J/\psi \to X\gamma \to \mu^+\mu^-\gamma$ as functions of the $X$ boson mass $m_X$ for the scalar and pseudoscalar $X$ bosons.}\label{psidwsps}
	\end{center}
\end{figure}

With the help of Eq. (\ref{dfG}), one can estimate the decay widths $\Gamma_{sig}$ of the signal process $J/\psi\to X\gamma \to \mu^+\mu^-\gamma$ as functions of the $X$ boson mass $m_X$ for both the scalar and pseudoscalar $X$ boson cases, as displayed in Fig. \ref{psidwsps}.
Here $\eta_c\eta_{\mu}=\xi_c\xi_{\mu}=10^{-6}$ is adopted, and the energy of emitting photon $E_{\gamma}$ is constrained in the region of $E_{\gamma}>100$ MeV.
And the total decay widths of $X$ boson $\Gamma_X$ = 10 eV, 100 eV and 1 keV are also adopted as three proper parameter options.
It is found that we have different line shapes for the decay widths of pseudoscalar and scalar $X$ bosons. 

It is worth noting that, in this spin-0 hypothesis, we can distinguish between the signal (Fig. \ref{psimmg} (c, d)) and the background (Fig. \ref{psimmg} (a, b)) by identifying where the emitting photon comes from. 
So the $J/\psi\to X\gamma \to \mu^+\mu^-\gamma$ process would be an ideal decay chain for the search of a spin-0 $X$ boson. 
According to Eq.s (\ref{sdw-1}, \ref{sdw-2}), we assume reasonably $\Gamma(X \to \mu^+\mu^-)\approx 1\sim10$ eV by setting $\eta_\mu \sim \xi_\mu \sim 10^{-3}$, then we obtain its decay length is $0.02\sim1~\mu$m for 270 MeV$<m_X<$ 2.8 GeV in the rest frame of $J/\psi$. 
Thus the secondary vertex reconstruction of the decay chain may help identifying the signal over the background.

\begin{figure}[htbp]
	\begin{center}
		\includegraphics[scale=0.85]{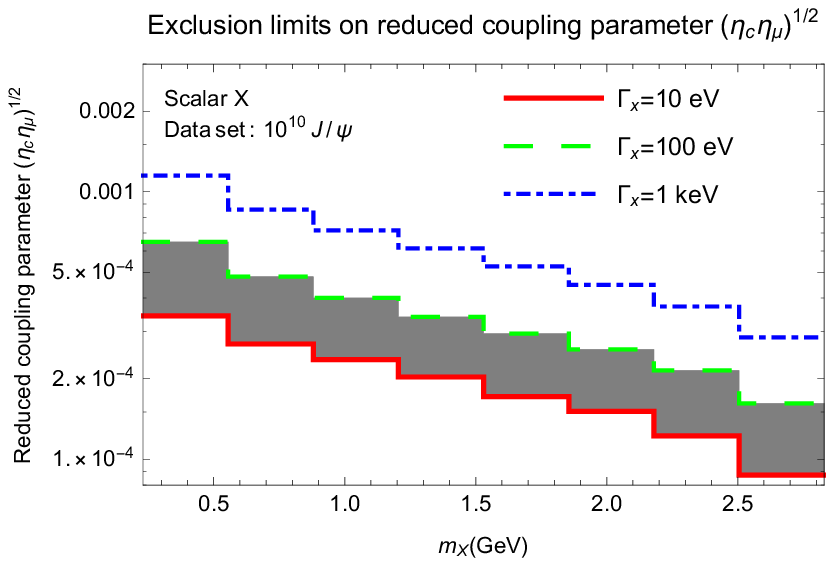}
		\includegraphics[scale=0.85]{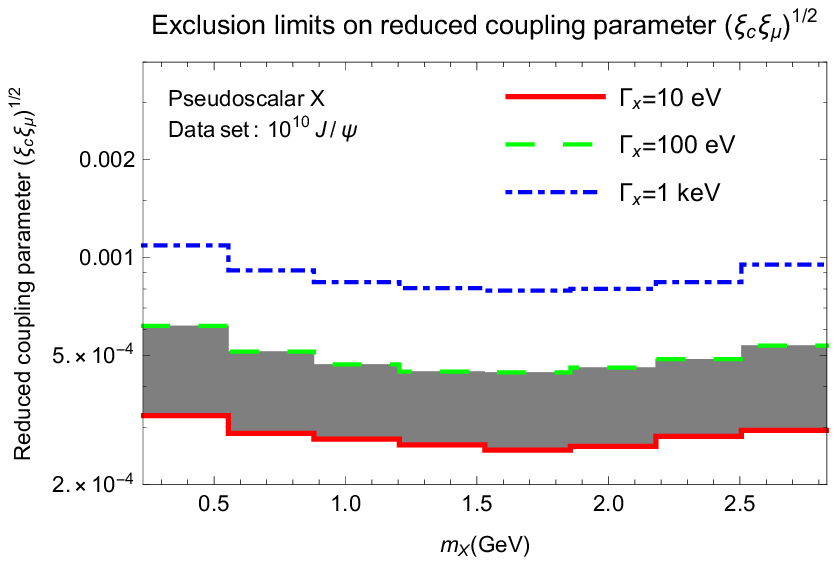}
		\caption{The exclusion limits on the reduced coupling strength parameters $(\eta_c\eta_{\mu})^{1/2}$ and $(\xi_c\xi_{\mu})^{1/2}$ as functions of  $m_X$ for the scalar and pseudoscalar $X$ respectively.}\label{psiexsps}
	\end{center}
\end{figure}

For the exclusion limits on the reduced coupling strength parameters as functions of $m_X$, the parameter spaces for the scalar and pseudoscalar $X$ bosons are presented in Fig. \ref{psiexsps}.
In these figures, the constrain on the energy of the emitting photon $E_{\gamma}>100$ MeV is adopted.
It is found that, with the 10 billion $J/\psi$ events, the upper limit on the parameter $(\eta_c\eta_{\mu})^{1/2}$ varies between $6\times10^{-4}$ and $1\times10^{-4}$ for 10 eV$<\Gamma_X<100$ eV depending on $m_X$. 
However, for the pseudoscalar case, the upper limit of $(\xi_c\xi_{\mu})^{1/2}$ stays around $4\times10^{-4}$ under the same conditions.
If the reduced coupling strength parameters were $\sqrt{\eta_c\eta_{\mu}}=\sqrt{\xi_c\xi_{\mu}}=10^{-3}$ as we adopted in Fig. \ref{psidwsps}, we believe that the scalar/pseudoscalar $X$ bosons may be found or excluded using the $\mu^+\mu^-$ mass spectrum of the $J/\psi \to \mu^+\mu^-\gamma$ process with the $10^{10}$ $J/\psi$ data set.

\subsection{Axial-vector Hypothesis}

Due to the fact that $J/\psi$ cannot decay into a dark photon $X$ and a SM photon, we will consider the $Z^0$-like $X$ hypothesis in this subsection.
By adopting the ``vector minus axial-vector" interaction Lagrangian of Eq. (\ref{lag}), one can obtain the decay width of the $J/\psi \to X + \gamma$ process,
\bqa
\Gamma(J/\psi\to X\gamma)=\frac{(\epsilon_a)^2 8 \pi \alpha ^2  \Psi^2 (16 \kappa^4+40 \kappa^2 +1)}{27 \kappa^4 (4 \kappa^2-1)m_X^2}, \label{G-av}
\eqa
where the variables have the same meanings as those in Eq.s (\ref{G-s/ps-1}, \ref{G-s/ps-2}).
Obviously, only the axial-vector current survives here.
One may also notice that Eq. (\ref{G-av}) resembles Eq. (\ref{G-s/ps-1}) of the scalar case. 
Given the total decay width $\Gamma_{J/\psi}=$ 92.9 keV and the 10 billion $J/\psi$ data set, we can estimate the events of this $Z^0$-like $X$ boson as $(0.65,~1.5,~6.2,~17.0) \times10^3$ for $m_X= (0.212,~1,~2,~2.5)$ GeV respectively by assuming $\epsilon_a=10^{-3}$, which are much less than the events produced in the $e^+e^- \to X+\gamma$ process when $m_X<2$ GeV, but then grow quickly with the increasement of $m_X$.

Next, we consider identifying the $Z^0$-like $X$ signals in the $J/\psi \to \mu^+\mu^-\gamma$ process, and we have four signal diagrams as displayed in Fig. \ref{psimmg}. 
Here, we divide them into two parts: Fig. \ref{psimmg} (a,~b) which have mutual interference with the background diagrams, and Fig. \ref{psimmg} (c,~d) or the $X$ resonance diagrams in $\mu^+\mu^-$ invariant mass spectrum. 
Then we calculate the differential decay width $d\Gamma/(d s_1d s_2)$ of those two parts and their cross terms, which have been sorted in the appendix. 
Here we also adopt the Breit-Wigner form for the $X$ propagators in Fig. \ref{psimmg} (c,~d). 
In addition, the differential decay width $d\Gamma/(d s_1d s_2)$ of the background process ($\gamma^*$-propagated Fig. \ref{psimmg} (a, b)) is also presented in the appendix.

\begin{figure}[htbp]
\begin{center}
\includegraphics[scale=0.85]{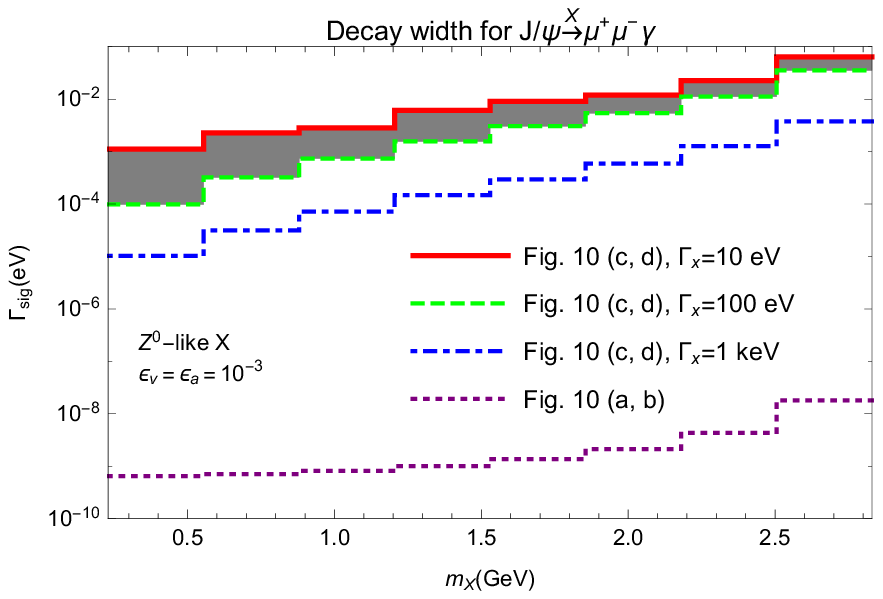}
\includegraphics[scale=0.85]{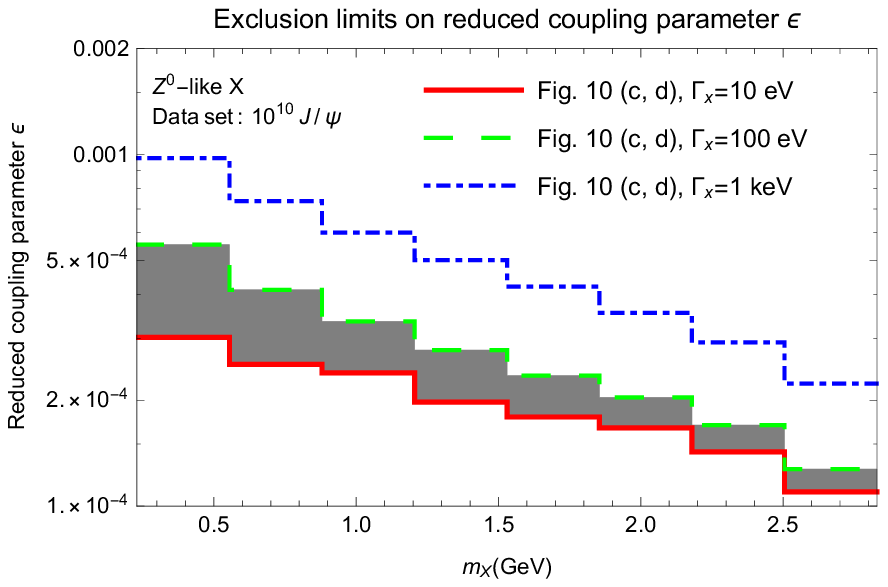}
\caption{The decay width $\Gamma_{sig}$ of the signal process and the exclusion limit on the reduced coupling strength parameter ($\epsilon_v=\epsilon_a=\epsilon$) as functions of the $X$ boson mass $m_X$ for $Z^0$-like $X$ boson.}\label{psidwav}
\end{center}
\end{figure}

With the help of differential Eq.s (\ref{dw-av-1}$\sim$\ref{dw-av-4}), we evaluate the decay width $\Gamma_{sig}$ of the signal process as a function of the $X$ boson mass $m_X$ in Fig. \ref{psidwav}.
We find that the Fig. \ref{psimmg} (c,~d) dominate the signal process for the $Z^0$-like interaction hypothesis. 
And the decay width of $X$-propagated Fig. \ref{psimmg} (a,~b) is $10^{-9}\sim10^{-8}$ eV depending on $m_X$. 
Moreover, for the decay width of $\gamma^*$-propagated Fig. \ref{psimmg} (a, b) (the background), we obtain $\Gamma_{bac}=1.5\times10^{-3}\Gamma_{J/\psi}=0.14$ keV with $E_{\gamma}>100$ MeV, which is absent in the PDG \cite{Tanabashi:2018oca}.
In Fig. \ref{psidwav}, we also present the parameter space of the reduced coupling strength ($\epsilon_v=\epsilon_a=\epsilon$) as a function of the $X$ boson mass $m_X$.
It is found that, with the 10 billion $J/\psi$ events, the upper limit on the parameter $\epsilon$ is placed in the range between $5\times10^{-4}$ and $1\times10^{-4}$ for 1 eV$<\Gamma_X<$ 100 eV depending on $m_X$.
In these plots, the energy of the emitting photon $E_{\gamma}$ is constrained in the region of $E_{\gamma}>100$ MeV.

The BESIII experiment searched for the light exotic $X$ particle twice in the $J/\psi \to \mu^+\mu^-\gamma$ process.
In 2011, they set 90\% confidence level upper limits on the product branching fraction $Br(J/\psi \to X\gamma)\times Br(X\to\mu^+\mu^-)$$<2.1\times10^{-5}$ \cite{Ablikim:2011es}.
In 2015, the upper limits on the product branching fraction were improved by a factor of five, below $5\times10^{-6}$ \cite{Ablikim:2015voa}.
In our estimation, assuming $\Gamma_X=10$ eV, the upper limits of the branching fraction $Br(J/\psi\to X\gamma\to\mu^+\mu^-\gamma)$ are estimated to be $2\times10^{-6}$, $2\times10^{-8}$ and $7\times10^{-7}$ for the scalar, pseudoscalar and $Z^0$-like $X$ hypotheses respectively, which are consistent with the experimental results.

\section{Summary and Outlook}

With the idea that non-SM gauge bosons imply new interactions beyond SM, in this work we investigate the possibility of searching for the new interaction mediator $X$ boson in the BESIII experiment.
We analyze the $X$ direct production in electron-positron collision and also indirect production in $J/\psi$ decay.
Four typical hypotheses for the $X$ boson, {\it i.e.} the scalar, pseudoscalar, $Z^0$-like and dark photon, prevailing in the literature are explored.
A systematic study on the production/decay properties of the $X$ boson, the product/decay chains $e^+e^-\to X\gamma \to e^+e^-\gamma$ and $J/\psi \to X\gamma\to\mu^+\mu^-\gamma$, and the exclusion limits on the reduced coupling strength parameters at BESIII are presented.

According to our analyses, some conclusions can be drawn here:
\begin{description}
  \item[(i)] For either spin-1 or spin-0 hypothesis, the total decay width $\Gamma_X$ of $X$ boson ranges roughly from 1 eV to 100 eV depending on $m_X$. One may obtain the bigger decay width when taking the $X$ to dark sector decay into consideration.
  \item[(ii)] For $X$ production process of $e^+e^- \xrightarrow{X} e^+e^-\gamma_{ISR}$, the $X$ resonant diagrams in $M_{ee}$ spectrum always dominate the cross section. And the contributions of vector and axial-vector currents are of the equal importance for $m_X>>m_e$ under the equal coupling strength assumption. It is found that the $Z^0$-like $X$ events produced in $e^+e^- \to X\gamma$ are more than those with $X$ being dark photon or spin-0 boson, and can reach $(10^{9}\sim10^{10})\times\epsilon^2$ per year depending on $m_X$. With tens of $\text{fb}^{-1}$ data set, we also found that the reduced coupling strength parameters ($\epsilon,~\eta_e,~\xi_e$) are placed in the range between $10^{-3}$ and $10^{-4}$ depending on $m_X$ for 1 eV$<\Gamma_X<$100 eV. These suggest that the $Z^0$-like boson $X$ signal might be found in the product chain $e^+e^- \to X\gamma$, $X \to e^+e^-$, or excluded, in the present run of BESIII.
  \item[(iii)]For $X$ production in the $J/\psi\to X\gamma,~X\to\mu^+\mu^-$ decay chain, the decay widths of $X$ being the scalar and $Z^0$-like bosons increase when the mass $m_X$ grows, while the line shape of $X$ being the pseudoscalar boson looks flat. Reasonably assuming 10 eV$<\Gamma_X<100$ eV, the coupling strength parameters ($\epsilon$, $\sqrt{\eta_c\eta_\mu}$) lie in the range of $5\times10^{-4}\sim1\times10^{-4}$ depending on $m_X$, while $\sqrt{\xi_c\xi_\mu}$ stays around $4\times10^{-4}$. We also find that it is an ideal channel to explore the spin-0 $X$ hypothesis, since one can distinguish the signal from the background by identifying the ISR photon. With the data set of $10^{10}$ $J/\psi$ events, one can obtain $O(10^3)$ scalar/$Z^0$-like $X$ bosons when setting the reduced Yukawa coupling parameters $\sim10^{-3}$, which is also within the reach of BESIII.
\end{description}

\section*{Appendix: Differential decay widths of $J/\psi(p) \xrightarrow[]{X/\gamma^*} \mu^+(k_1)+\mu^-(k_2)+\gamma(k_3)$}

\begin{center}
    \noindent{\it{1. Background process}}
\end{center}

We firstly introduce some notations, $s_1=(k_1+k_2)^2$, $s_2=(k_2+k_3)^2$, $\Psi^2$ is the squared wave function at the origin of $J/\psi$ and $\Gamma_X$ is the total decay width of the $X$ boson.

Then for the background process $J/\psi \xrightarrow[]{\gamma^*} \mu^+\mu^-\gamma$ (Fig. \ref{psimmg} (a, b)), we obtain its differential decay width,
\bqa
\frac{d\Gamma}{d s_1 d s_2}=\frac{\alpha^3 \Psi^2}{768 m_c^4}\times |M_{bg}|^2,
\eqa
with
\bqa
&&|M_{bg}|^2=
\frac{128}{9 m_c^2 (m_{\mu}^2-s_2)^2 (-4 m_c^2-m_{\mu}^2+s_1+s_2)^2}\nonumber\\
&&\times \big(-s_2^2 (-8 s_1 (2 m_c^2+m_{\mu}^2)+12 (2 m_c^2+m_{\mu}^2)^2+3 s_1^2)+s_2 (4 m_c^2+2 m_{\mu}^2-s_1) (4 (2 m_c^2+m_{\mu}^2)^2+s_1^2)\nonumber\\
&&+4 s_2^3 (4 m_c^2+2 m_{\mu}^2-s_1)+m_{\mu}^2 (-192 m_c^6+16 m_c^4 (5 s_1-7 m_{\mu}^2)-4 m_c^2 (4 m_{\mu}^4-8 m_{\mu}^2 s_1+3 s_1^2)\nonumber\\
&&-2 m_{\mu}^6-3 m_{\mu}^2 s_1^2+s_1^3)-2 s_2^4\big). 
\eqa

\begin{center}
    \noindent{\it {2. $Z^0$-like $X$ case}}
\end{center}

The differential decay width of the signal process $J/\psi \xrightarrow[]{X} \mu^+\mu^-\gamma$ for the $Z^0$-like $X$ boson (Fig. \ref{psimmg} (a$\sim$d)),
\bqa
\frac{d\Gamma}{d s_1 d s_2}=\frac{\alpha^3 \Psi^2}{768 m_c^4}\times(|M_1|^2+|M_{2}|^2+|M_3|^2),\label{dw-av-1}
\eqa
where $|M_1|^2$ is the squared amplitudes of the signal diagrams of Fig. \ref{psimmg} (a,~b), $|M_3|^2$ stands for those of Fig. \ref{psimmg} (c,~d), and $|M_2|^2$ is their cross terms. And their explicit expressions are
\bqa
&&|M_1|^2=
\frac{36 \epsilon_v^4 m_c^4 |M_{bg}|^2}{(m_X^2-4 m_c^2)^2}+\frac{256 \epsilon_v^2\epsilon_a^2}{(m_X^2-4 m_c^2)^2 (m_{\mu}^2-s_2)^2 (-4 m_c^2-m_{\mu}^2+s_1+s_2)^2}\nonumber\\
&&\times\big(128 m_c^8 (s_2-3 m_{\mu}^2)+32 m_c^6 (3 m_{\mu}^4+m_{\mu}^2 (5 s_1+8 s_2)-s_2 (s_1+3 s_2))\nonumber\\
&&-8 m_c^4 (6 m_{\mu}^6-2 m_{\mu}^4 (s_1+8 s_2)+m_{\mu}^2 (3 s_1^2+22 s_1 s_2+14 s_2^2)-s_2 (s_1+2 s_2)^2)\nonumber\\
&&-2 m_c^2 (2 m_{\mu}^8-8 m_{\mu}^6 (2 s_1+s_2)+3 m_{\mu}^4 (3 s_1^2+12 s_1 s_2+4 s_2^2)-m_{\mu}^2 (s_1^3+20 s_1^2 s_2+24 s_1 s_2^2+8 s_2^3)\nonumber\\
&&+s_2 (s_1+s_2) (s_1^2+2 s_1 s_2+2 s_2^2))-m_{\mu}^2 s_1^2 (m_{\mu}^2-s_2) (m_{\mu}^2-s_1-s_2)\big),\label{dw-av-2}
\eqa
\bqa
|M_{2}|^2=\frac{1024 \epsilon_v^2\epsilon_a^2 (s_1-m_X^2) (16 m_c^4 (3 m_{\mu}^2-2 s_1)+16 m_c^2 m_{\mu}^2 s_1+m_{\mu}^2 s_1^2)}{3 (4 m_c^2-m_X^2) (m_{\mu}^2-s_2) (\Gamma_X ^2 m_X^2+(m_X^2-s_1)^2) (4 m_c^2+m_{\mu}^2-s_1-s_2)},\label{dw-av-3}
\eqa
\bqa
&&|M_{3}|^2=
\frac{1024 \epsilon_a^2}{9 (s_1-4 m_c^2)^2 (\Gamma_X^2 m_X^2+(m_X^2-s_1)^2)}\nonumber\\
&& \times \big(\epsilon_a^2\big(16 m_c^4 (s_1-2 m_{\mu}^2)-8 m_c^2 s_1 (13 m_{\mu}^2-3 s_1+s_2)+s_1 (2 m_{\mu}^4-4 m_{\mu}^2 (s_1+s_2)+s_1^2+2 s_1 s_2+2 s_2^2)\big)\nonumber\\
&& +\epsilon_v^2\big(16 m_c^4 (2 m_{\mu}^2+s_1)+8 m_c^2 s_1 (7 m_{\mu}^2+3 s_1-s_2)+s_1 (2 (m_{\mu}^4-2 m_{\mu}^2 s_2+s_2 (s_1+s_2))+s_1^2)\big)\big). \label{dw-av-4}
\eqa

\vspace{.7cm}
{\bf Acknowledgments}

We thank Xiao-Rui Lyu for the discussion on BESIII detector. This work was supported in part by the Ministry of Science and Technology of the Peoples' Republic of China(2015CB856703); by the Strategic Priority Research Program of the Chinese Academy of Sciences, Grant No.XDB23030100; by the National Natural Science Foundation of China(NSFC) under the Grants 11375200 and 11635009; and by the Fundamental Research Funds of Shandong University (2019GN038).



\begin{thebibliography}{99}

\bibitem{Tanabashi:2018oca}
  M.~Tanabashi {\it et al.} [Particle Data Group],
  Phys.\ Rev.\ D {\bf 98}, no. 3, 030001 (2018).

\bibitem{Aaboud:2018zba}
  M.~Aaboud {\it et al.} [ATLAS Collaboration],
  arXiv:1801.08769 [hep-ex].
\bibitem{Aaboud:2018jux}
  M.~Aaboud {\it et al.} [ATLAS Collaboration],
  [arXiv:1807.10473 [hep-ex]].
\bibitem{Kalsi:2018dgi}
  A.~K.~Kalsi, J.~B.~Singh and V.~Bhatnagar,
  Springer Proc.\ Phys.\  {\bf 203}, 443 (2018).
\bibitem{CMS:2018jxa}
  CMS Collaboration [CMS Collaboration],
  CMS-PAS-B2G-18-001


\bibitem{Ahmed:2016ift}
  H.~Ahmed,
  AIP Conf.\ Proc.\  {\bf 1742}, 030001 (2016).
\bibitem{TheBelle:2015mwa}
  I.~Jaegle [Belle Collaboration],
  Phys.\ Rev.\ Lett.\  {\bf 114}, no. 21, 211801 (2015),
  [arXiv:1502.00084 [hep-ex]].
  \bibitem{Ablikim:2011es}
  M.~Ablikim {\it et al.} [BESIII Collaboration],
  Phys.\ Rev.\ D {\bf 85}, 092012 (2012),
  [arXiv:1111.2112 [hep-ex]].
\bibitem{Ablikim:2015voa}
  M.~Ablikim [BESIII Collaboration],
  Phys.\ Rev.\ D {\bf 93}, no. 5, 052005 (2016),
  [arXiv:1510.01641 [hep-ex]].

\bibitem{Banerjee:2018vgk}
  D.~Banerjee {\it et al.} [NA64 Collaboration],
  Phys.\ Rev.\ Lett.\  {\bf 120}, no. 23, 231802 (2018),
  [arXiv:1803.07748 [hep-ex]]
\bibitem{Lees:2017lec}
  J.~P.~Lees {\it et al.} [BaBar Collaboration],
  Phys.\ Rev.\ Lett.\  {\bf 119}, no. 13, 131804 (2017),
  [arXiv:1702.03327 [hep-ex]].
\bibitem{Aaij:2017rft}
  R.~Aaij {\it et al.} [LHCb Collaboration],
  Phys.\ Rev.\ Lett.\  {\bf 120}, no. 6, 061801 (2018),
  [arXiv:1710.02867 [hep-ex]].
\bibitem{Bloise:2015wfb}
  C.~Bloise [KLOE-2 Collaboration],
  Acta Phys.\ Polon.\ A {\bf 127}, 1565 (2015).
\bibitem{Agakishiev:2013fwl}
  G.~Agakishiev {\it et al.} [HADES Collaboration],
  Phys.\ Lett.\ B {\bf 731}, 265 (2014),
  [arXiv:1311.0216 [hep-ex]].

\bibitem{Raggi:2018doy}
  M.~Raggi,
  EPJ Web Conf.\  {\bf 179}, 01020 (2018).
\bibitem{Wojtsekhowski:2017ijn}
  B.~Wojtsekhowski {\it et al.},
  JINST {\bf 13}, no. 02, P02021 (2018),
  [arXiv:1708.07901 [hep-ex]].
\bibitem{Corliss:2017tms}
  R.~Corliss [DarkLight Collaboration],
  Nucl.\ Instrum.\ Meth.\ A {\bf 865}, 125 (2017).


\bibitem{Bauer:2017ris}
  M.~Bauer, M.~Neubert and A.~Thamm,
  JHEP {\bf 1712}, 044 (2017),
  [arXiv:1708.00443 [hep-ph]].
\bibitem{Bauer:2018uxu}
  M.~Bauer, M.~Heiles, M.~Neubert and A.~Thamm,
  arXiv:1808.10323 [hep-ph].

\bibitem{Wang:2018ycf}
  D.~Wang,
  Int.\ J.\ Mod.\ Phys.\ Conf.\ Ser.\  {\bf 46}, 1860046 (2018).
\bibitem{Ablikim:2018bhf}
  M.~Ablikim {\it et al.} [BESIII Collaboration],
  arXiv:1809.00635 [hep-ex].
\bibitem{Ablikim:2017aab}
  M.~Ablikim {\it et al.} [BESIII Collaboration],
  Phys.\ Lett.\ B {\bf 774}, 252 (2017),
  [arXiv:1705.04265 [hep-ex]]

\bibitem{Holdom:1986eq}
  B.~Holdom,
  Phys.\ Lett.\ B {\bf 178}, 65 (1986).
\bibitem{Dienes:1996zr}
  K.~R.~Dienes, C.~F.~Kolda and J.~March-Russell,
  Nucl.\ Phys.\ B {\bf 492}, 104 (1997),
  [hep-ph/9610479].
\bibitem{Kahn:2016vjr}
  Y.~Kahn, G.~Krnjaic, S.~Mishra-Sharma and T.~M.~P.~Tait,
  JHEP {\bf 1705}, 002 (2017),
  [arXiv:1609.09072 [hep-ph]].
\bibitem{Ismail:2016tod}
  A.~Ismail, W.~Y.~Keung, K.~H.~Tsao and J.~Unwin,
  Nucl.\ Phys.\ B {\bf 918}, 220 (2017),
  [arXiv:1609.02188 [hep-ph]]
\bibitem{Ellwanger:2009dp}
  U.~Ellwanger, C.~Hugonie and A.~M.~Teixeira,
  Phys.\ Rept.\  {\bf 496}, 1 (2010),
  [arXiv:0910.1785 [hep-ph]].

\bibitem{Fayet:1989mq}
  P.~Fayet,
  Phys.\ Lett.\ B {\bf 227}, 127 (1989).
\bibitem{Das:1999hn}
  P.~Das and P.~Jain,
  Phys.\ Rev.\ D {\bf 62}, 075001 (2000),
  [hep-ph/9903432].
\bibitem{Emam:2007dy}
  W.~Emam and S.~Khalil,
  Eur.\ Phys.\ J.\ C {\bf 52}, 625 (2007),
  [arXiv:0704.1395 [hep-ph]].
\bibitem{Kyae:2013hda}
  B.~Kyae and C.~S.~Shin,
  JHEP {\bf 1306}, 102 (2013),
  [arXiv:1303.6703 [hep-ph]].
\bibitem{Dong:2016gxl}
  P.~V.~Dong, D.~T.~Huong, D.~V.~Loi, N.~T.~Nhuan and N.~T.~K.~Ngan,
  Phys.\ Rev.\ D {\bf 95}, no. 7, 075034 (2017),
  [arXiv:1609.03444 [hep-ph]]
\bibitem{Alonso:2017uky}
  R.~Alonso, P.~Cox, C.~Han and T.~T.~Yanagida,
  Phys.\ Lett.\ B {\bf 774}, 643 (2017),
  [arXiv:1705.03858 [hep-ph]].
\bibitem{Nomura:2017lsn}
  T.~Nomura and H.~Okada,
  JHEP {\bf 1801}, 099 (2018),
  [arXiv:1710.10028 [hep-ph]].

\bibitem{Krasznahorkay:2015iga}
  A.~J.~Krasznahorkay {\it et al.},
  Phys.\ Rev.\ Lett.\  {\bf 116}, no. 4, 042501 (2016),
  [arXiv:1504.01527 [nucl-ex]].
\bibitem{Feng:2016jff}
  J.~L.~Feng, B.~Fornal, I.~Galon, S.~Gardner, J.~Smolinsky, T.~M.~P.~Tait and P.~Tanedo,
  Phys.\ Rev.\ Lett.\  {\bf 117}, no. 7, 071803 (2016),
  [arXiv:1604.07411 [hep-ph]].
\bibitem{Gu:2016ege}
  P.~H.~Gu and X.~G.~He,
  Nucl.\ Phys.\ B {\bf 919} (2017) 209,
  [arXiv:1606.05171 [hep-ph]].

\bibitem{Liang:2016ffe}
  Y.~Liang, L.~B.~Chen and C.~F.~Qiao,
  Chin.\ Phys.\ C {\bf 41}, no. 6, 063105 (2017),
  [arXiv:1607.08309 [hep-ph]].
\bibitem{Jia:2017iyc}
  L.~B.~Jia,
  Eur.\ Phys.\ J.\ C {\bf 78}, no. 2, 112 (2018),
  [arXiv:1710.03906 [hep-ph]].

\bibitem{Fayet:1981rp}
  P.~Fayet and M.~Mezard,
  Phys.\ Lett.\  {\bf 104B}, 226 (1981).
\bibitem{Batell:2009yf}
  B.~Batell, M.~Pospelov and A.~Ritz,
  Phys.\ Rev.\ D {\bf 79}, 115008 (2009),
  [arXiv:0903.0363 [hep-ph]].
\bibitem{Bjorken:2009mm}
  J.~D.~Bjorken, R.~Essig, P.~Schuster and N.~Toro,
  Phys.\ Rev.\ D {\bf 80}, 075018 (2009),
  [arXiv:0906.0580 [hep-ph]].
\bibitem{Li:2009wz}
  H.~B.~Li and T.~Luo,
  Phys.\ Lett.\ B {\bf 686}, 249 (2010),
  [arXiv:0911.2067 [hep-ph]].
\bibitem{Alikhanov:2017cpy}
  I.~Alikhanov and E.~A.~Paschos,
  Phys.\ Rev.\ D {\bf 97}, no. 11, 115004 (2018),
  [arXiv:1710.10131 [hep-ph]]
\bibitem{Araki:2017wyg}
  T.~Araki, S.~Hoshino, T.~Ota, J.~Sato and T.~Shimomura,
  Phys.\ Rev.\ D {\bf 95}, no. 5, 055006 (2017),
  [arXiv:1702.01497 [hep-ph]].
\bibitem{Alioli:2017nzr}
  S.~Alioli, M.~Farina, D.~Pappadopulo and J.~T.~Ruderman,
  Phys.\ Rev.\ Lett.\  {\bf 120}, no. 10, 101801 (2018),
  [arXiv:1712.02347 [hep-ph]].
\bibitem{He:2017zzr}
  M.~He, X.~G.~He, C.~K.~Huang and G.~Li,
  JHEP {\bf 1803}, 139 (2018),
  [arXiv:1712.09095 [hep-ph]].

\bibitem{Jiang:2018uhs}
  J.~Jiang, L.~B.~Chen, Y.~Liang and C.~F.~Qiao,
  Eur.\ Phys.\ J.\ C {\bf 78}, no. 6, 456 (2018),
  [arXiv:1607.03970 [hep-ph]].


\bibitem{Han:2005mu}
  T.~Han,
  hep-ph/0508097.
  
\bibitem{Ablikim:2009aa}
  M.~Ablikim {\it et al.} [BESIII Collaboration],
  Nucl.\ Instrum.\ Meth.\ A {\bf 614}, 345 (2010),
  [arXiv:0911.4960 [physics.ins-det]].

\bibitem{Asner:2008nq} 
D.~M.~Asner {\it et al.},
Int.\ J.\ Mod.\ Phys.\ A {\bf 24}, S1 (2009),
[arXiv:0809.1869 [hep-ex]].



\bibitem{Braaten:2002fi}
  E.~Braaten and J.~Lee,
  Phys.\ Rev.\ D {\bf 67}, 054007 (2003),
  Erratum: [Phys.\ Rev.\ D {\bf 72}, 099901 (2005)],
  [hep-ph/0211085].







\end{thebibliography}
\end{document}